\documentclass{moriond}

\usepackage{mathtools}
\usepackage{lineno}

\def\Journal#1#2#3#4{{#1} {\bf #2}, #3 (#4)}

\def\NIMA{{\em Nucl. Instrum. Methods} A}

\def\PLB{{\em Phys. Lett.}  B}

\def\PRD{{\em Phys. Rev.} D}

\def\be{\begin{equation}}
\def\ee{\end{equation}}
\def\bea{\begin{eqnarray}}
\def\eea{\end{eqnarray}}

\newcommand\myeq{\stackrel{~\mathclap{\textrm{\mbox{SM}}}}{=}~}
\newcommand\maxeq{\stackrel{~\mathclap{\textrm{\mbox{max}}}}{=}~}

\newcommand{\Photo}{}

\begin{document}

\vspace*{4cm}
\title{$\tau$ and low multiplicity decays at Belle and Belle II}

\author{ Philipp Horak on behalf of the Belle II collaboration}

\address{HEPHY Vienna, Austrian Academy of Sciences}

\maketitle
\abstracts{
We present recent measurements of $\tau$ physics at the Belle II experiment at SuperKEKB.  Measurements include a test of $e-\mu$ lepton flavor universality in $\tau$ decays, a search for the lepton flavor violating $\tau \rightarrow \mu \mu \mu$ decay and a precision measurement of the $\tau$ lepton invariant mass. In addition, we present a measurement of the low multiplicity decay cross section $\sigma(e^+ e^- \rightarrow \pi^+ \pi^- \pi^0)$, a key input in the theoretical prediction of the anomalous magnetic moment of the muon.}
\section{Introduction}
$B$ factories, such as the Belle II experiment \cite{tdr} at the SuperKEKB collider \cite{superkekb}, collide anti-symmetric $e^+ e^-$ beams at center-of-mass energies of $10.58$ GeV. The cross section of $\tau \overline \tau$ production in these decays $\sigma(e^+ e^- \rightarrow  \tau \overline\tau)  \simeq 0.9$ nb is of similar order of magnitude as the $b \overline b$ production cross section $\sigma(e^+ e^- \rightarrow \Upsilon(4S) \rightarrow b \overline b) \simeq 1.1$ nb, allowing a large sample of $\tau$ decays to be recorded in the Belle II data. \\
The $\tau$ lepton is the heaviest known lepton with a mass of $1.777$ GeV, and therefore the only lepton that can decay hadronically. Studying the $\tau$ lepton and its decays is interesting in multiple aspects. The properties of the $\tau$ such as its invariant mass $m_\tau$ are fundamental standard model parameters that have to be obtained from measurement, and precise knowledge is necessary as input in searches for new physics and in other Standard Model precision measurements. Furthermore, fundamental properties of the SM, such as the coupling of the $W$ boson to the different light lepton flavors $\frac{g_e}{g_\mu}$, can be studied precisely in $\tau$ decays. Finally, some NP scenarios predict enhanced contributions of lepton flavor violating decays (LFV) such as $\tau \rightarrow \mu \mu \mu$. Such decays are allowed in the SM with neutrino mixing at rates of $\mathcal{O}(10^{-50})$, but enhanced significantly in NP scenarios. The Belle II sample of $\tau$ decays can be used to set limits on LFV branching ratios, constraining the NP scenarios.\\
The Belle II detector offers an excellent environment for studying $\tau$ decays. The precise knowledge of the $e^+ e^-$ initial state and the near-hermetic coverage allows probing processes with invisible contributions from neutrinos with high precision. The total data sample collected up to now corresponds to $424$ fb$^{-1}$, of which $362$ fb$^{-1}$ were collected at the $\Upsilon(4S)$ resonance energy, and contains $\sim4\cdot10^8$ $\tau$ pairs available for analysis.\\
In addition to $\tau$ physics, the detector properties of Belle II are suitable for low-multiplicity decays. The $ee \rightarrow \mathrm{hadrons}$ cross sections are an important input in the context of the $(g-2)_\mu$ anomaly~\cite{gm2ti}. The theoretical prediction of the muon anomalous magnetic moment $a_\mu$ depends on the precise knowledge of the center-of-mass energy dependent ratio 
\begin{equation}
    \label{eq:hadronicR}
    R(s) = \frac{\sigma(e^+ e^- \rightarrow \mathrm{hadrons})}{\sigma(e^+ e^- \rightarrow \mu^+ \mu^-)}.
\end{equation}
The second largest contribution to this ratio stems from the $e^+ e^- \rightarrow \pi^+ \pi^- \pi^0$ decay. By analyzing $e^+e^-$ events with an initial-state radiation photon, the cross section can be measured at \mbox{Belle II} for different initial state center-of-mass energies.
\section{$\tau$ event reconstruction at Belle II}
Events involving $\tau$ pairs are characterized by low track multiplicity and large missing energy. The decay products can be separated into two hemispheres defined by the plane perpendicular to the thrust axis
\begin{equation}
    V_{\mathrm{thrust}} ~\maxeq~ \frac{\sum_i |\vec{p}^{~\mathrm{CM}}_i~\cdot~ \hat{n}_{\mathrm{thrust}}|}{\sum_i |\vec{p}^{~\mathrm{CM}}_i|},
\end{equation}
where $i$ runs over charged final state particles in the event. \\
The hemisphere with the decay of interest is labelled \textit{signal} and the other hemisphere is labelled \textit{tag}. To suppress backgrounds, a certain topology can be required on the tag side. The common choices are \textit{1-prong}, \textit{i.e.} one charged track on the tag side or \textit{3-prong}, \textit{i.e.} three charged tracks on the tag side. 
\section{Lepton flavor universality in $\tau$ decays}
The SM predictions for the ratio of the $\tau$ branching ratios and the coupling of the $W$ boson to light leptons $e$ and $\mu$ are 
\begin{equation}
R_\mu = \frac{\mathcal{B}(\tau^- \rightarrow \mu^- \overline{\nu}_\mu \nu_\tau)}{\mathcal{B}(\tau^- \rightarrow e^- \overline{\nu}_e \nu_\tau)} ~\myeq~ 0.9726.
\end{equation}
and 
\begin{equation}
\left(\frac{g_{\mu}}{g_{e}}\right)_{\tau}=\sqrt{R_\mu\frac{f(m_{e}^{2}/m_{\tau}^{2})}{f(m_{\mu}^{2}/m_{\tau}^{2})}}    ~\myeq~ 1 
\end{equation}
The deviation from unity in $R_\mu$ follows from the mass difference of the involved leptons \cite{leptonmass}. A wide range of SM extensions couple differently to each of the three lepton flavor generations \cite{lfunp1}$^{,~}$\cite{lfunp2}, motivating a search for lepton flavor universality (LFU) violation in $\tau$ decays.\\
We report a measurement of LFU in $\tau$ decays using a data set corresponding to  $362$ fb$^{-1}$. Events are selected with a 1-prong tag side with at least one $\pi^0$ present. To suppress backgrounds, a selection is placed on a neural network classifier trained on simulation. The obtained combined sample for $e$ and $\mu$ shows a purity of $94\%$ at $9.6\%$ signal efficiency, with the main residual backgrounds coming from $e^+ e^- \rightarrow \tau^+ \tau^-$ events with misidentified leptons or wrongly reconstructed tag sides. The signal yields are extracted in a simultaneous maximum likelihood fit for both $e$ and $\mu$ modes in $p_\ell$ using \texttt{pyhf} \cite{pyhf}. Systematic uncertainties are included in the fit as nuisance parameters, allowing shape and normalization variations of the fit templates. \\
The ratio of branching ratio leads to large cancellation of systematic uncertainties, with the biggest exception being the impact of lepton identification. Lepton identification correction factors and uncertainties are derived from calibration samples such as $ee \rightarrow ee(\gamma)$ and $ee \rightarrow \mu \mu \gamma$ for correctly reconstructed leptons and $K_S^0 \rightarrow \pi^+ \pi^-$ and $\tau \rightarrow \pi \pi \pi \nu$ for misidentified leptons. Lepton identification makes up for $0.32 \%$ of the total systematic uncertainty of $0.37\%$, with the next biggest contribution being $0.10 \%$ from the trigger efficiency.\\
We measure $R_\mu = 0.9675 \pm 0.0007_{\mathrm{stat.}} \pm 0.0036_{\mathrm{syst.}}$ or, converted to couplings, $\left(\frac{g_{\mu}}{g_{e}}\right)_{\tau}=0.9974 \pm 0.0019$. The result is consistent within $1.4\sigma$ with the SM expectation and is the world's most precise measurement of $e-\mu$ universality in $\tau$ decays yet.
\section{Search for $\tau \rightarrow \mu \mu \mu$}
The $\tau \rightarrow \mu \mu \mu$ mode is an excellent candidate to search for LFV, with the fully leptonic final state allowing for significant suppression of backgrounds. Its branching ratio is predicted to be $10^{-53} - 10^{-56}$ in the SM with neutrino mixing \cite{3mu1}. NP scenarios such as the seesaw mechanism or various SUSY models significantly enhance this contribution to branching ratios of $10^{-8} - 10^{-10}$ \cite{3mu2}$^{,~}$\cite{3mu3}.\\
We report a Belle II measurement using a data sample corresponding to 362 fb$^{-1}$. We use an inclusive approach on the tag side, combining both 1-prong and 3-prong decays. Backgrounds are suppressed by using a combination of rectangular selections and a BDT classifier.
We obtain a signal efficiency of $(20.42\pm0.06)\%$ at $0.5^{+1.4}_{-0.5}$ expected background events. The signal window is defined by the invariant mass of the three $\mu$ system $M_{3\mu}$ and the energy difference $\Delta E_{3\mu} = E_{\tau, sig} - E_{beam}$. After box opening, one event is found in the signal region, consistent with the expectation. This allows us to set an upper limit at $90\%$ CL of $\mathcal{B}(\tau \rightarrow \mu \mu \mu) < 1.9 \times 10^{-8}$, resulting in the world's most stringent constraint on this decay mode yet.
\section{$\tau$ mass measurement}
The $\tau$ mass is a fundamental parameter of the SM and an important input for, e.g., the $\tau$ LFU measurement. We present a determination using the pseudo-mass method with a data set corresponding to 190 fb$^{-1}$.  We reconstruct $\tau^- \rightarrow \pi^- \pi^+ \pi^- \nu_\tau$ events and construct the pseudo-mass     $M_\mathrm{min} = \sqrt{M^2_{3\pi} + 2 (\sqrt{s} /2 - E^*_{3\pi}) (E^*_{3\pi} - p^*_{3\pi})}$. This quantity shows a rapid drop off at the $\tau$ mass, with small tails from detector resolution effects and initial-state radiation. The limiting factor in this measurement is maintaining high precision in the reconstruction of track momenta of the final state particles, and the precise determination of the beam energy. Track momenta are calibrated and validated in control samples to produce polar-angle dependent correction factors, limiting the impact of the systematic uncertainty to $\sim 70$ keV. The beam energy spread is measured and corrected for using $B$ decays, with the uncertainty of the correction totalling $\sim 60$ keV. The fit to the $M_\mathrm{min}$ end-point yields $m_\tau = 1777.09 \pm 0.08_{\mathrm{stat.}} \pm 0.11_{\mathrm{syst.}}\mathrm{MeV/c^2}$, the world's most precise measurement of the $\tau$ mass up to date \cite{taumass}.  
\section{Differential cross section of $e^+ e^- \rightarrow \pi^+ \pi^- \pi^0$ decays}
An observed tension between theoretical predictions and measurements of the anomalous magnetic moment of the muon $a_\mu = \frac{(g-2)_\mu}{2}$ has been a heavily discussed topic in particle physics in recent years. Originally reported as a discrepancy of up to $\sim 5\sigma$ \cite{gm2ti}$^{,~}$\cite{gm2fermilab}, recent results from the BMW lattice group and the CMD-3 collaboration reduce the tension to as low as $\sim 1 \sigma$ when used in the fit. The contributions to the $a_\mu$ prediction split up into $a_\mu = a^{\mathrm{EW}}_\mu + a^{\mathrm{QED}}_\mu + a^{\mathrm{QCD}}_\mu$ and furthermore $a^{\mathrm{QCD}}_\mu = a^{\mathrm{HVP}}_\mu + a^{\mathrm{HLbL}}_\mu$, with a hadron vacuum polarization and a light-by-light term. The largest contribution to the overall uncertainty stems from the HVP term, which further depends on the hadronic $R$ ratio given in Eq. \ref{eq:hadronicR}. As a function of the center-of-mass energy $\sqrt{s}$, $R$ is dominated by contributions around the $\omega$ and $\phi$ resonances at $\sim0.8$ and $\sim1$ GeV respectively. All $ee \rightarrow \mathrm{hadrons}$ cross sections contribute to the ratio, with the main contributions coming from $e^+e^- \rightarrow \pi^+ \pi^-$ and $e^+e^- \rightarrow \pi^+ \pi^- \pi^0$ decays. The current knowledge of the $e^+e^- \rightarrow \pi^+ \pi^- \pi^0$ cross section makes up for $\sim 15\%$ of the total error budget of $a^{\mathrm{HVP}}_\mu$. Previous measurements of the cross section show tensions of up to $2-3\%$ between experiments, particularly around the $\omega$ resonance \cite{gm2snd1}$^{-}$\cite{gm2babar2}. \\
We present a Belle II measurement of $e^+e^- \rightarrow \pi^+ \pi^- \pi^0$ in a data set corresponding to $191$ fb$^{-1}$. By explicitly reconstructing an initial state radiated photon, the available center-of-mass energy of the $\pi^+ \pi^- \pi^0$ system can be varied in a wide range of 0.62 GeV to 3.5 GeV. Among other rectangular selections, a kinematic fit is performed by constraining the sum of $\pi^+ \pi^- \pi^0 \gamma_{ISR}$ momenta to the $e^+ e^-$ beam momentum, allowing effective suppression of backgrounds. The main remaining backgrounds stem from $e^+ e^- \rightarrow \pi^+ \pi^- \pi^0 \pi^0 \gamma$, $e^+e^- \rightarrow K^+ K^- \pi^0 \gamma$ and $e^+ e^- \rightarrow q \overline{q}$ decays. The backgrounds are each studied and calibrated in dedicated control samples.\\
A limiting factor in the analysis is the effect of the $\pi^0$ reconstruction efficiency on the total signal efficiency. A partial reconstruction approach is performed with events at the $\omega$ resonance by computing 
\begin{equation}
    \varepsilon_{\pi^0} = \frac{N(\textrm{Full reconstruction of }\gamma_{ISR}~ \pi^+ \pi^- \pi^0)}{N(\textrm{Partial reconstruction of }\gamma_{ISR}~ \pi^+ \pi^-)}.
\end{equation}
From the data-to-simulation ratio of $\varepsilon_{\pi^0}$, a correction factor with associated uncertainty is derived for the $\pi^0$ efficiency, resulting in a systematic uncertainty of $1.0\%$. An excess with respect to the \texttt{PHOKHARA} generator observed at BABAR of events with a single additional energetic photon, and the absence of events with two additional ISR photons in the generator affect the signal efficiency and contribute a total systematic uncertainty of $1.2\%$.\\
The signal yields are extracted by fitting the $\pi^0 \rightarrow \gamma \gamma$ $M_{\gamma\gamma}$ spectrum in bins of the three-pion invariant mass $M_{3\pi}$. The contribution to the leading-order HVP term in $a_\mu$ is obtained by integrating over the measured cross section from 0.62 to 1.8 GeV to obtain \cite{gm2belle2} $a^{3\pi}_{\mu} = (48.91 \pm 0.23_\mathrm{stat.} \pm 1.07_\mathrm{syst.}) \times 10^{-10}$. The measured value of $a^{3\pi}_{\mu}$ is $6.9\%$ or $2.5\sigma$ higher than the value observed by BABAR \cite{gm2babar2}, and $6.5\%$ higher than the global fit \cite{gm2global}.  
\section{Summary}
We presented recent $\tau$ and low multiplicity physics results with subsets of the total data set collected so far by the Belle II experiment. In addition to the large data sample available, continuously improving analysis techniques and reduced systematic uncertainties allow us to achieve world-leading precision in measurements of fundamental SM properties and parameters, as well as searches for physics beyond the SM.

\end{document}